\documentclass[aps,amsmath,amssymb,preprintnumbers,superscriptaddress,twocolumn,floatfix,superscriptaddress
]{revtex4-2}

\usepackage{hyperref}% add hypertext capabilities
\usepackage{xcolor}

\newcommand{\be}{\begin{equation}}
\newcommand{\ee}{\end{equation}}
\newcommand{\beq}{\begin{equation}}
\newcommand{\eeq}{\end{equation}}
\newcommand{\ba}{\begin{aligned}}
\newcommand{\ea}{\end{aligned}}

\newcommand{\bea}{\begin{eqnarray}}
\newcommand{\eea}{\end{eqnarray}}

\newcommand{\cT}{\mathcal{T}}

\newcommand{\cN}{\mathcal{N}}

\newcommand{\cV}{\mathcal{V}}

\newcommand{\cM}{\mathcal M}

\newcommand\bi{\begin{itemize}}
\newcommand\ei{\end{itemize}}

\def\Im{\mathop{\mathrm{Im}}\nolimits}

\def\Tr{\mathop{\mathrm{Tr}}\nolimits}

\def\unit{{1\kern-.65ex {\rm l}}}
\def\1{{1\kern-.65ex {\rm l}}}

\def\ii{{\rm i}}
\usepackage{graphicx}% Include figure files
\usepackage{dcolumn}% Align table columns on decimal point
\usepackage{bm}% bold math
\begin{document}
 \preprint{UPR-1335-T}

\title{Smooth String Vacua in a Gravitationally Non-perturbative Regime}

\author{Mirjam Cveti\v{c}}
\affiliation{
 Department of Physics and Astronomy, University of Pennsylvania,  Philadelphia, PA 19104, USA
}%
\affiliation{
Department of Mathematics, University of Pennsylvania,  Philadelphia, PA 19104, USA}
\affiliation{ Center for Applied Mathematics and Theoretical Physics, University of Maribor, Maribor, Slovenia}
\author{Max Wiesner}
\affiliation{%
 II. Institut f\"ur Theoretische Physik, Universit\"at Hamburg, Notkestrasse 9, 22607 Hamburg, Germany
}%

\date{\today}

\begin{abstract}
The strong coupling regime of four-dimensional N=2 supersymmetric vacua of the heterotic string is analyzed from a dual domain wall perspective. Using modular invariance, we compute a closed form for the non-perturbative corrections to the supersymmetric domain wall equations, which enables a quantitative study of gravitational strong coupling regimes. A strong coupling singularity for the hidden Ho\v{r}ava-Witten 9-brane is resolved, and the domain wall interpolates between the visible 9-brane and a supersymmetric Anti-de Sitter vacuum, thereby realizing a variant of the Randall-Sundrum model. 
\end{abstract}

\maketitle

%\tableofcontents

\section*{Introduction}
Quantum Field Theory provides a consistent and remarkably successful framework for describing fundamental particle interactions, yet its strongly coupled regimes---such as confinement in Quantum Chromodynamics and strong-field phenomena in Quantum Electrodynamics---remain inaccessible to perturbative techniques. A consistent quantum theory of gravity poses an even greater challenge. While String Theory stands as a leading candidate, it is primarily defined through a perturbative expansion and thus offers limited control over non-perturbative and strong-coupling effects. Nevertheless, assuming spacetime to have sufficiently many spatial dimensions and backgrounds preserving enough supersymmetry, certain strong coupling regimes of String Theory can be reliably described due to strong-weak coupling dualities, implying that the strong coupling regime is accessible using a dual perturbative description. 

For theories in four dimensions (4d) that preserve eight or less supercharges, the web of dualities does not cover all strong coupling regimes. Whereas certain strongly coupled field theory sectors can be described using geometric engineering techniques associated with string compactifications, \emph{gravitational} strong-coupling regimes in 4d theories of quantum gravity are relatively unexplored. For this reason, most attempts to relate String Theory to the observed Universe either by constructing explicit string vacua or in the context of the Swampland program~\cite{Vafa:2005ui} (see \cite{Carta:2016ynn,Palti:2019pca,Grana:2021zvf,vanBeest:2021lhn,Agmon:2022thq} for reviews) rely on a perturbative description of String Theory. However, by restricting to the perturbative ``String Theory Landscape'', we remain blind to potentially interesting gravitational physics associated with the Landscape of strongly coupled 4d string vacua.

In this letter, we report new results on the nature of gravitationally strongly coupled string vacua in 4d. By using a domain wall (DW) as a probe for strong coupling physics, we obtain crucial insights into the nature of the strongly coupled phases of compactifications of Heterotic String Theory to 4d. Concretely, we consider the strong-coupling phase of the Heterotic $E_8\times E_8$ String Theory compactified on a Calabi--Yau twofold (K3) times a two-torus ($T^2$) preserving $\cN=2$ supersymmetry in 4d. Classically, the strong coupling regime corresponds to Ho\v{r}ava--Witten (HW) theory, i.e., M-theory on a K3$\times T^2 \times S^1/\mathbb{Z}_2$ with two 9-branes located at the end of the interval~\cite{Horava:1995qa}. In this setup, the heterotic string coupling $g_h$ is identified with the length, $\rho$, of the interval $S^1/\mathbb{Z}_2$. For Calabi--Yau compactifications of the heterotic string, it was already noticed in~\cite{Witten:1996mz} that, for generic gauge bundles, one of the $E_8$ gauge theories (or what remains of the $E_8$ gauge group after Higgsing) becomes strongly coupled for finite values of $2\pi \rho$ inducing a singularity in the moduli space. However, in 4d, perturbative and non-perturbative effects have to be taken into account to fully capture the strong-coupling physics. In~\cite{Cvetic:2024wsj} a step in this direction was achieved by using the results of~\cite{Alexandrov:2023hiv} to study the leading non-perturbative corrections in a dual DW picture.
 
Similarly, the 4d $\cN=2$ setups studied here can be described as a DW solution in 5d $\cN=2$ gauged supergravity arising from M-theory on K3$\times T^2$. Crucially, the parent theory exhibits a modular $SL(2,\mathbb{Z})$ symmetry for the modulus that obtains a non-trivial profile in the DW solution. Exploiting this modular symmetry we determine the full non-perturbative corrections to the DW equations of motion paralleling the study of cosmic string solutions in theories exhibiting modular symmetry in~\cite{Greene:1989ya}. We demonstrate explicitly that in this case the strong coupling singularity in the moduli space is resolved, allowing the DW solution to extend infinitely in the fifth dimension. At infinity, the DW flows to a supersymmetric Anti-de Sitter (AdS) vacuum realized at the self-dual point of the modular symmetry. The AdS vacuum thus effectively replaces one of the HW 9-branes. The strong-coupling regime of the heterotic string on K3$\times T^2$ thus corresponds to a thick DW in an AdS$_5$ space which can be interpreted as a top-down realization of a thickened variant of the Randall--Sundrum 2 model~\cite{Randall:1999vf}. Importantly, even though the DW extends to infinity, the graviton is confined to a finite region in the fifth dimension~\cite{Cvetic:2008gu} such that the strong coupling regime of the thick DW does not correspond to a decompactification limit in which gravity propagates in five macroscopic dimensions.

\section*{Classical Domain Wall Solutions and Strong Coupling}\label{sec:7d}

Consider the heterotic $E_8\times E_8$ string compactified on K3$\times T^2$ with gauge embedding $(12-n,12+n)$, $n\neq0$, in the K3, leading to a 4d $\cN=2$ theory of supergravity in the low-energy limit. We are interested in the strong coupling regime corresponding to $g_h\to \infty$ at fixed 4d heterotic dilaton
\begin{equation}\label{4ddilaton}
  \text{Re}\,S_{\rm 4d} = \frac{\text{vol}(K3\times T^2) M_{\rm het}^6}{g_h^2}\,.
 \end{equation}
Classically, this limit is best described using the HW M-theory dual of the heterotic $E_8\times E_8$ string~\cite{Horava:1995qa}, i.e., M-theory compactified on $K3\times T^2\times S^1/\mathbb{Z}_2$. There are two parameters of relevance: the length, $2\pi \rho$, of the interval and the overall volume of K3$\times T^2$
\begin{equation}
V = \text{vol}(K3\times T^2)M_{11}^6\,,
\end{equation}
with $M_{11}$ the eleven-dimensional Planck scale. The limit $g_h\to \infty$ for the heterotic string at constant $\text{Re}\,S_{\rm 4d}$ corresponds to $\rho\to \infty$ at constant $V$. Classically, this limit is thus a decompactification limit to M-theory compactified on K3$\times T^2$ with gravity propagating in five dimensions.

The 5d parent theory preserves 16 supercharges and has a supergravity and 21 matter multiplets. The massless scalars span the coset space~\cite{Romans:1986er,Witten:1995ex,Aspinwall:1996mn}
\begin{equation}
 \cM_{5d,\cN=2} = \Gamma_{5,21} \Big\backslash \frac{O(5,21)}{O(5)\times O(21)} \times \mathbb{R}^+\,. 
\end{equation} 
The physical moduli parametrizing $\cM_{5d,\cN=2}$ are best understood in the dual Type IIA compactification on K3$\times S^1$: The $\mathbb{R}^+$ factor is parametrized by the 5d dilaton 
\begin{equation}
 e^{-2\phi_5} = \frac{1}{g_s^2} (\text{vol}\,(K3) M_s^4) (R_{S^1} M_s)\,,
\end{equation} 
where $g_s$ is the Type IIA string coupling, $M_s$ the string scale and $R_{S^1}$ is the $S^1$ radius. All other moduli reside in the first factor in $\cM_{5d,\cN=2}$. In particular, this applies to all other combinations of $g_s, \text{vol}(K3) M_s^4$ and $R_{S^1}M_s$ such as 
\begin{equation}
 V = \frac{1}{g_s}( \text{vol}(K3) M_s^4)\, (R_{S^1} M_s) = g_s e^{-2\phi_5}\,. 
\end{equation} 
Using the dictionary between M-theory and Type IIA String Theory $
 M_s = g_s^{1/3} M_{11}$ and $(R_{11} M_{11})^{3/2} = g_s$ with $R_{11}$ the radius of the M-theory circle, we identify
\begin{equation}
 V= (\text{vol}(K3) M_{11}^4) (\text{vol}(T^2) M_{11}^2) \,.
\end{equation} 
The scalar $V$ is the real part of a complex scalar
\begin{equation}\label{def:T}
 T = V + \ii \theta_5= V+ \int_{K3\times T^2} C_6\,,
\end{equation} 
with $C_6$ the M-theory six-form. Inside $\cM_{{\rm 5d},\cN=2}$, $T$ parametrizes a sub-manifold
\begin{equation}\begin{aligned}\label{eq:ModulispaceT}
 &\left[SL(2,\mathbb{Z})\Big\backslash \frac{SL(2,\mathbb{R})}{U(1)}\right]_T \subset \Gamma_{5,21} \Big\backslash \frac{O(5,21)}{O(5)\times O(21)}\,,
\end{aligned}\end{equation} 
to which we restrict in the following.

By compactifying this 5d $\cN=2$ supergravity theory on an interval $S^1/\mathbb{Z}_2$ we obtain HW theory. Due to the orbifold action, there are two 9-branes located at the end of the interval wrapping the K3. We will refer to the 9-brane located at $y=0$ as $\mathbf{9}_+$ and to the one at $y=2\pi \rho$ as $\mathbf{9}_-$ with $y$ the coordinate along the interval.  Both 9-branes host a (possibly Higgsed) gauge group contained in $E_8$ for which the gauge kinetic function $f_{\mathbf{9}_\pm}$ is classically given by $f_{\mathbf{9}_\pm} = 2\pi T$. In the M-theory picture, the heterotic gauge bundle with instanton number $(12-n,12+n)$ maps to $G_4$ flux localized in the 9-branes at $y=0$ and $y=2\pi \rho$ given by~\cite{Lukas:1998yy,Lukas:1998tt}
\begin{equation}
 Q_{G_4}^{\pm} = \Tr F^{(\pm)}\wedge F^{(\pm)}- \frac12  \text{tr} R\wedge R \,,
\end{equation} 
where the first term is the class of the heterotic gauge bundle in the two 9-branes.
The second term is the  second Chern class of K3 such that $Q_{\rm G4}^\pm = \pm n$ \footnote{In HW language, the heterotic anomaly cancellation corresponds classically to the condition $Q^+_{\rm G4}+ Q^-_{\rm G4}=0$.}. The fluxes $Q_{G_4}^{\pm}$ effectively induce a gauging of the 5d supergravity yielding an effective potential $\cV_{\rm eff}$ for the modulus $V$ given by~\cite{Skenderis:1999mm}
\begin{equation}
\cV_{\rm eff} = 6 e^{K_{\rm 5d}}\left[3|D_T W_{\rm 5d,eff}|^2 -4|W_{\rm 5d,eff}|^2\right)\,,
\end{equation} 
where $W_{\rm 5d,eff}$ is the flux-induced superpotential 
\begin{equation} \label{eq:supclass}
 W_{\rm 5d, eff}=  Q_{G_4}^+\,,
\end{equation} 
such that
\begin{equation}\label{eq:cVclass}
 \cV_{\rm eff, cl} = - 6 e^{K_{\rm 5d}(T,\bar{T})} n^2 \,.
\end{equation} 
Here, we used that classically the K\"ahler potential for the modulus $T$ is given by~\footnote{Strictly speaking, there is no Kahler potential in 5d theories of supergravity but on the slice of the moduli space that we are interested in, we can define a function from which the moduli space metric is derived, see~\cite{Cvetic:2024wsj} for details.}
\begin{equation}\label{eq:Kclass} 
 K_{\rm 5d, cl.} = - \log (T+\bar {T}) \,. 
\end{equation} 
If $Q_{G_4}^+=0$, we have $W_{\rm 5d,eff}=0$ and the product manifold ${\rm K3}\times T^2\times S^1/\mathbb{Z}_2\times \mathbb{R}^{1,3}$ is a solution to the supergravity equations of motion (EOM). Instead, for $Q_{G_4}^+\neq 0$, the product manifold is not a solution to the EOM. To obtain a solution, the modulus $T$ has to vary along the interval $S^1/\mathbb{Z}_2$ and the spacetime metric has to be non-trivially warped. To obtain a supersymmetric solution to the EOM, we make an ansatz for the 5d metric 
\begin{equation}\label{metricansatz}
 {\rm d}s^2_{5} = e^{2a} \mathrm{d}x^\mu \mathrm{d}x^\nu \eta_{\mu \nu} + e^{8a} {\rm d}y^2 \,,\quad \mu,\nu=0,\dots,3\,,
\end{equation}
where $x^\mu$ are coordinates of $\mathbb{R}^{1,3}$ and $y$ the coordinate along $S^1/\mathbb{Z}_2$. Using this ansatz, we can identify solutions that preserve supersymmetry. In this case, the EOM are equivalent to the first order equations~\cite{Cvetic:1992bf,Cvetic:1994ya,Cvetic:1996vr,Skenderis:1999mm}\footnote{The EOM for co-dimension one objects are the same irrespective of the total dimension of spacetime. Here, we use 4d $\cN=1$ language as in~\cite{Cvetic:1992bf,Cvetic:1994ya,Cvetic:1996vr} even though this language is not the most natural in five dimensions.}
\begin{equation}\label{eq:DWBPS}
\begin{aligned}
\partial_y a(y) &= \mp \frac14 e^{4a} |W| e^{K/2} \,,\\
\partial_y T &= \pm \frac34e^{K/2} |W| K^{T\bar T} \frac{D_{\bar T} \overline{W}}{\overline{W}} \,,
\end{aligned}
\end{equation}
which determine how the warp factor $a$ and the modulus $T$ vary along the interval $S^1/\mathbb{Z}_2$. The resulting geometry can be viewed as a DW solution for which the modulus $T$, i.e., the volume of K3$\times T^2$, varies over the interval. A solution to these equations is given by~\cite{Lukas:1998yy,Lukas:1998tt} 
\begin{equation}\label{solution1}
a_0 e^{6a} = \text{Re}\,T(y)=V_0 H(y)^3 \,.
\end{equation}
$H(y)$ is a harmonic function satisfying 
\begin{equation}
\partial_y^2 H(y) = -\frac{2\sqrt{2}}{3} \left(Q_{G_4}^+ \delta(y) + Q_{G_4}^{-} \delta(y-2\pi \rho)\right)\,.
\end{equation}
Here, we used that the fluxes localized on the $\mathbf{9}_\pm$-brane effectively act as $\delta$-sources that trigger a non-trivial profile $T(y)$. One then finds
\begin{equation}
 H(y) =-\frac{2\sqrt{2}}{3} Q_{G_4}^+ |y| + c_0\,.
\end{equation} 
For $Q_{G_4}^+\neq0$ we hence obtain a warped spacetime metric and a non-trivial profile for $T$ along the interval. The resulting setup can be viewed as a DW solution interpolating between the $\mathbf{9}_+$- and $\mathbf{9}_-$-brane. The integration constant $c_0>0$ determines the value of $\text{Re}\,T$ at the location of $\mathbf{9}_+$. Crucially, $T(y_*)=0$ for 
\begin{equation}\label{eq:ystar}
 y_* = \frac{3c_0}{2\sqrt{2} Q_{G_4}^+} >0\,. 
\end{equation} 
By \eqref{solution1}, the warp factor diverges if we position the brane $\mathbf{9}_-$ at $y=y_*$ signaling the presence of an end-of-the-world brane (see also~\cite{Buratti:2021fiv} for a recent discussion). Moreover, since $\text{Re}\,T(2\pi \rho)$ measures the (inverse) gauge coupling of the gauge theory on $\mathbf{9}_-$, this gauge theory becomes strongly coupled if we choose $2\pi \rho=y_*$ (see also~\cite{Witten:1996mz}). Recall that duality to the Heterotic String identifies the interval length, $\rho$, with the heterotic string coupling. Since the DW solution can be studied for any $\rho$, we use this dual DW picture to study the Heterotic String at finite string coupling in the following.

\section*{Quantum Corrections to Domain Wall Solutions}\label{sec:QCDW}
To compute the effect of non-perturbative corrections on the DW equations, we make use of the $SL(2,\mathbb{Z})_T$ modular symmetry appearing in~\eqref{eq:ModulispaceT}~\footnote{See\cite{Greene:1989ya} for a similar approach to obtain cosmic string solutions in 4d.}. 
Consider the effective K\"ahler covariant superpotential
\begin{equation}\label{def:G}
 G_{\rm eff} = K_{\rm 5d} + \log |W_{\rm 5d,eff}|^2\,. 
\end{equation} 
This combination of K\"ahler and superpotential has to be invariant under the action of the modular group $SL(2,\mathbb{Z})_T$. Together with the boundary conditions for $T\to \infty$, namely
\begin{equation}\label{Gasympt}
 G_{\rm eff} \stackrel{T\to \infty}{\longrightarrow} -\log(T+\bar{T}) + \log n^2\,,
\end{equation} 
modular invariance can be used to determine the non-perturbative form of $G_{\rm eff}$. The classical K\"ahler potential \eqref{eq:Kclass} has modular weight -2 under general transformations 
\begin{equation}
 i T \to \frac{i aT+b}{icT+d}\,,\qquad \begin{pmatrix}a&b\\ c & d \end{pmatrix} \in {\rm SL}(2,\mathbb{Z})\,.  
\end{equation} 
As in~\cite{Greene:1989ya}, this transformation property is compensated by suitable factors of the Dedekind eta function $\eta(iT)$. Imposing also the boundary condition at $T=\infty$, the invariant form of $G_{\rm eff}$ takes the form~\footnote{Since the Kahler metric on the moduli space is inherited from the hyperbolic metric on the upper half space, all corrections to $G_{\rm eff}$ have to be of the form of a Kahler transformation thereby excluding the addition of modular weight $(-2)$ to the classical $(T+\bar{T})$-term appearing in the Kahler potential.} 
\begin{equation}\begin{aligned}\label{Ginv}
 G_{\rm eff} =& -\log\left[(T+\bar{T}) \eta(iT)^2 \eta(-i\bar{T})^2 \right] \\&+ \log |n j(iT)^{-1/12}|^2\,,
\end{aligned}\end{equation} 
where the asymptotic behavior in~\eqref{Gasympt} is ensured by the power of the $SL(2,\mathbb{Z})$-invariant $j$-function~\footnote{The above is a minimal choice for the dependence of $G_{\rm eff}$ on the $j$-function. In principle, one could add to $G_{\rm eff}$ the quotient of two polynomials in $j$ of the same degree without affecting the asymptotics. This could, however, induce additional, unphysical zeros/poles for $G_{\rm eff}$ within the fundamental domain. It would be interesting to confirm the above form of $G_{\rm eff}$ by explicitly computing M5-brane instanton corrections to superpotential in 5d gauged supergravity.}. For large $\text{Re}\,T$, the corrections can be interpreted as M5-brane instantons with action $S_{M5} = 2\pi T $. As a result of the corrections, the scalar potential of the 5d gauged supergravity now reads 
\begin{equation}\begin{aligned}\label{eq:cVcorr} 
 &\cV_{\rm eff}(T,\bar{T}) = \frac{6n^2}{(T+\bar{T}) |\eta(iT)|^4 |j(iT)|^{1/6}}\\
 &\times \left[3(T+\bar T)^2 \left|\frac{1}{2\pi}\hat{G}_2(T,\bar T) -\frac{1}{12} \frac{\partial_T j(iT)}{j(iT)}\right|^2-4\right]\,,
\end{aligned}\end{equation} 
where $\hat G_2(T,\bar{T})$ is the non-holomorphic Eisenstein series 
\begin{equation}
\hat{G}_2(T,\bar{T}) = -4 \pi \partial_T \log \eta -  \frac{2\pi}{T +\bar{T}} \,. 
\end{equation}
For $T\to \infty$, the potential scales as ~\eqref{eq:cVclass} and diverges for $iT=(-1)^{1/3}$. At $T=1$ the derivative of $G_{\rm eff}$ vanishes 
\begin{equation*}
 \partial_T G_{\rm eff}\bigg|_{T=1} =  \left[\frac{n}{2\pi}\hat{G}_2(T,\bar T) -\frac{n}{12} \frac{\partial_T j(iT)}{j(iT)}\right]\bigg|_{T=1} =0 \,,
\end{equation*} 
implying that there is a supersymmetric AdS vacuum at $T=1$ with cosmological constant
\begin{equation}\label{LambdaAdS}
 \Lambda_{\rm AdS}  = -\frac{12 n^2 }{ |\eta(i)|^4 |j(i)|^{1/6}} < 0\,. 
\end{equation}
We now return to the DW EOM~\eqref{eq:DWBPS}. The point $T=1$ being the only supersymmetric vacuum indicates that this point is a universal attractor for the DW EOM. Including the quantum corrections, the DW solution differs from the classical solution discussed around~\eqref{solution1}. Let us fix $T(y=0)\equiv T_0 \gg 1$ such that the DW solution maps the vicinity of $\mathbf{9}_+$ to a regime in moduli space where all instanton effects are suppressed. The brane $\mathbf{9}_+$ acting as a $\delta$-source, the solution to the DW equations~\eqref{eq:DWBPS} is locally of the form~\eqref{solution1}. If we make the interval large enough, the effects of the instantons on~\eqref{eq:DWBPS} become important. Sufficiently far away from $y=0$, the profile of the DW solution is determined entirely by the first-order equations with the corrected expressions for $K_{5d}$ and $W_{\rm 5d,eff}$. Since $T=1$ is the universal attractor, the DW solution approaches the vacuum at $T=1$ for $y\to \infty$, i.e., $T(y) = 1 + \dots$ and
\begin{equation}
 e^{-4a} \stackrel{y\to \infty}{\longrightarrow}  \left(\frac{n^2}{(T+\bar{T}) |\eta(iT)|^4 |j(iT)|^{1/6}}\right)^{1/2} y +\tilde{c}\,,\end{equation} 
for some constant $\tilde{c}$. Inserting this solution in the metric~\eqref{metricansatz} we indeed realize an AdS$_5$ geometry for $y\to \infty$. If we choose $\text{Im}\,T_0=0$, the path described by the DW in the fundamental $SL(2,\mathbb{Z})$ domain is a straight line along the real axis ending at $T=1$. If instead we turn on the axion along $\mathbf{9}_+$, i.e. $\Im\,T_0\neq 0$, the DW solution is not a straight line in the fundamental domain but eventually bends towards the universal attractor at $\text{Re}\,T=1$ and $\text{Im}\,T=0$. 
\section*{Phases of 4d $\cN=2$ heterotic M-theory}\label{sec:4dphases}
From the perspective of the Heterotic $E_8\times E_8$ String on K3$\times T^2$, we consider a particular sector of the moduli space spanned by the complexified 4d dilaton~\eqref{4ddilaton} and overall volume modulus $\cT_{\rm het}$. As discussed in~\cite{Cvetic:2024wsj}, the heterotic variables can be translated into the parameters of the HW setup using 
\begin{equation}\label{eq:dictionary}
 \text{Re}\,T_0= \text{Re}\,S_{\rm 4d} \,,\qquad \rho = \frac{1}{(\text{Re}\,S_{\rm 4d})^{1/3}} \cT_{\rm het}\,,
\end{equation} 
where $T_0$ is the value of the complex $T$-field defined in \eqref{def:T} at $y=0$, i.e., the position of $\mathbf{9}_+$. We now discuss the interpretation of the strong coupling regime of the Heterotic String using the quantum corrected DW solution described in the previous section. \newline

\textbf{Strong-coupling Phase.}
In the classical heterotic moduli space, we can consider the limit $\cT_{\rm het}\gg \text{Re}\,S_{\rm het}$, i.e., the large volume limit while keeping the 4d dilaton finite. By~\eqref{eq:dictionary}, this limit corresponds to the regime $\rho\gg 1$. Comparing~\eqref{eq:dictionary} with \eqref{4ddilaton} we find $\rho = g_{\rm het}^{2/3} $, such that $\cT_{\rm het}\gg \text{Re}\,S_{\rm het}$ is the strong coupling regime for the 10d heterotic string coupling. Classically, one expects the strong coupling limit of the heterotic string to correspond to a decompactification limit to a theory in which the massless graviton propagates in five dimensions. However, for the 4d heterotic string, this is not necessarily the case since perturbative threshold corrections induce a strong coupling singularity for one of the perturbative heterotic gauge groups~\cite{Cvetic:2024wsj}. In HW theory this singularity corresponds to choosing $2\pi \rho=y_*$ which yields an end-of-the-world defect as $a(y)\to -\infty$~\cite{Witten:1996mz,Lukas:1998yy}. 

The non-perturbative instanton corrections to the DW solution smooth out the singularity in the warp factor and the moduli space extends to larger $\rho$. Classically, the length of the interval determines the region along which the graviton propagates in the fifth dimension. For large $\rho$, the DW solution can be viewed as smoothly connecting the HW solution valid for $y\ll y_*$ with a thick supergravity DW for $y\gg y_*$ asymptoting to the AdS$_5$ vacuum at $T=1$. For such a thick DW, the massless graviton is confined to a region corresponding to the thickness of the DW~\cite{Cvetic:2008gu}. The thickness of the DW can be estimated by the region in which the profile of $T(y)$ varies significantly before the non-perturbative corrections enforce it to asymptote to its critical value at $T=1$. As the non-perturbative corrections become important at $T\sim \mathcal{O}(1)$, using~\eqref{solution1} the thickness of the domain wall can be estimated to be $\mathcal{O}(c_0/n)$. 

Recall that in the weak-coupling phase, the fifth dimension is an interval obtained as a $\mathbb{Z}_2$ orbifold of a circle. For large $\rho$, the $\mathbf{9}_-$-brane is replaced by the supersymmetric AdS$_5$ vacuum. Still, the $\mathbb{Z}_2$ symmetry remains intact at the quantum level such that the supersymmetric AdS$_5$ vacuum is realized on both sides of the $\mathbf{9}_+$-brane in a $\mathbb{Z}_2$ symmetric way. The resulting setup is thus a variant of the model discussed in~\cite{Randall:1999vf}, commonly referred to as Randall--Sundrum Model 2. However, instead of an infinitely thin wall as in~\cite{Randall:1999vf}, here we have a domain wall of finite thickness of $\mathcal{O}(c_0/n)$. Therefore, gravity is not completely confined to the worldvolume of a brane as in~\cite{Randall:1999vf} but extends into a finite region of the fifth dimension similar to the discussion in~\cite{Cvetic:2008gu}. 

In summary, for finite $c_0$,  the region to which gravity is confined is compact, and our results have the following interpretation from the perspective of the Heterotic String: Perturbative effects remove the classical infinite distance limit $\rho\to \infty$, inducing a finite distance strong-coupling singularity. This singularity gets resolved by non-perturbative effects since the warp factor remains finite everywhere allowing for a smooth transition into a compact strong-coupling phase in which the perturbative  heterotic gauge theory associated with the $E_8$ factor realized in $\mathbf{9}_-$ is expected to confine.\newline

\textbf{Spectrum of States.}
From the analysis of the quantum-corrected DW solution, we infer that we can smoothly transition between the weakly coupled phase corresponding to the perturbative heterotic string and the strong coupling phase where the HW $\mathbf{9}_-$-brane is replaced by a supersymmetric AdS vacuum. In the weak coupling phase, the spectrum of states is determined by the excitations of the perturbative heterotic string.  Since one of the gauge groups $G\subset E_8$ becomes strongly coupled in the strong coupling phase, the light states associated in this phase can no longer be the perturbative string excitations electrically charged under the $E_8$. 

Instead, there is a monopole string that determines the strong coupling physics. At weak coupling, this string corresponds to the M5-brane wrapping the K3 factor localized in the brane $\mathbf{9}_-$ with tension~\footnote{The exponent accounts for the fact that $T$ is classically the overall volume of $K3\times T^2$.}
\begin{equation}\label{eq:tensionM5}
 \frac{T_{M5|\mathbf{9}_-}}{M_{\rm pl,4}^2} = \left(e^{ K_{\rm 5d}(T,\bar{T})} T \right)^{2/3}\bigg|_{y=2\pi \rho} \,,
 \end{equation} 
 We can trace this string into the strong coupling phase~\footnote{Even though the geometric realization of the string is not obvious in this regime, there is still a BPS string with tension given by the expression on the rhs of \eqref{eq:tensionM5} which now has to be evaluated at $y\to \infty$.}, where its minimal tension is
\begin{equation}
\frac{T_{M5|{\rm AdS}}}{M_{\rm pl,4}^2} =( e^{K_{\rm 5d}(T,\bar{T})}T)^{2/3}\bigg|_{y=\infty} = \frac{\Lambda_{\rm AdS}^{2/3}}{4\cdot 3^{2/3} n^{4/3}}\,.
\end{equation}
Here we used $T(y\to \infty)=1$ and inserted \eqref{LambdaAdS}. Accordingly, the minimal tension for this string is of the order of AdS$_5$ scale.

Whereas the weak coupling phase is characterized by the perturbative string excitations, the strong coupling phase is governed by a string with tension of order the 5d AdS scale. In particular, even though the DW solution extends to $y\to \infty$, there is no tower of KK modes becoming massless reflecting that the graviton is confined to a finite region in the fifth dimension. 

\section*{Conclusions}\label{sec:conclusions}
In this letter, we analyzed the strong coupling phase of 4d $\cN=2$ heterotic string vacua from a dual DW perspective. In this dual description, which we reviewed below~\eqref{eq:Kclass}, we were able to explicitly compute the non-perturbative quantum corrections by exploiting the modular invariance of the five-dimensional parent theory of quantum gravity. Our results show that the non-perturbative effects resolve a strong coupling singularity in the moduli space. Moreover, the closed form for the non-perturbative corrections allowed us to study the nature of the strong coupling phase of 4d $\cN=2$ heterotic string vacua. Concretely, the DW solution revealed that in the strong coupling phase, one of the HW 9-branes is replaced by a supersymmetric AdS$_5$ vacuum. We interpret this as the gravitational version of the brane-flux transition used in~\cite{Atiyah:2000zz} to describe confinement for gauge theories in the absence of gravity. Interestingly, the quantum-corrected DW solution provides a concrete top-down realization for the Randall--Sundrum 2 model~\cite{Randall:1999vf} with a thick DW embedded in AdS$_5$ and gravity confined to a finite region in the five-dimensional spacetime. 

It would be very interesting to extend this analysis to the setup with minimal supersymmetry studied in~\cite{Cvetic:2024wsj}. A key difference in this case is that the metric on the moduli space of the 5d parent theory itself receives (non-)perturbative corrections such that the effect of M5-brane instantons cannot be written as a K\"ahler transformation as in~\eqref{Ginv}. Still, it would be interesting to unveil the nature of the strong coupling phase of 4d $\cN=1$ compactifications of the heterotic string. Given the duality between the heterotic string on K3$\times T^2$ and Calabi--Yau threefold compactifications of Type II String Theory~\cite{Kachru:1995wm} (see also~\cite{Lee:2019oct,Friedrich:2025gvs}), it would further be interesting to understand the Type II dual of the strong coupling phase discussed here. Understanding the Type II dual of the heterotic strong coupling phase can also shed new light on strong coupling regimes in 4d $\cN=1$ F-theory~\cite{Wiesner:2022qys} that are dual to the heterotic strong coupling regimes studied in~\cite{Cvetic:2024wsj}.\newline 

We thank Cumrun Vafa for useful discussions and the Harvard Swampland Initiative and Harvard Physics Department for hospitality during the completion of this work. The work of MC is supported by DOE (HEP) Award DE-SC0013528, the Slovenian Research Agency (ARRS No. P1-0306) and Fay R. and Eugene L. Langberg Endowed Chair funds. The work of MW is supported in part by Deutsche Forschungsgemeinschaft under Germany’s Excellence Strategy EXC 2121 Quantum Universe 390833306 and through the Collaborative Research Center 1624 “Higher Structures, Moduli Spaces and Integrability.”

\bibliography{papers_Max}% Produces the bibliography via BibTeX.

\end{document}